# Exploring Structural and Electronic Properties of Topological Insulator/Graphene Nano-heterostructures

Valentina Gallardo [a], Barbara Arce [a], Francisco Muñoz [b], Rodolfo San Martín [a], Irina Zubritskaya [c], Paula Giraldo-Gallo [d], Hari Manoharan [c] and Carolina Parra *[e,f]

[a.] Departamento de Física, Universidad Técnica Federico Santa María, Avenida España 1680, Valparaíso, Chile.
[b.] Departamento de Física, Facultad de Ciencias, Universidad de Chile, 9170124 Santiago, Chile.
[c.] Department of Physics, Stanford University, Stanford, California 94305, United States.
[d.] Department of Physics, Universidad de Los Andes, Bogotá 111711, Colombia.
[e.] Departamento de Ingeniería Mecánica, Universidad Técnica Federico Santa María, Avenida España 1680, Valparaíso, Chile. E-mail: carolina.parra@usm.cl
[f.] Centro Científico Tecnológico de Valparaíso CCTVaL Universidad Técnica Federico Santa María, Valparaíso 2390123, Chile.

## Abstract

There is great interest in the study of topological insulator-based heterostructures due to expected emerging phenomena. However, a challenge of topological insulator (TI) research is the contribution of the bulk conduction to the TI surface states. Both strain engineering and thickness control routes, which have been proposed to compensate for bulk doping, can be accessed through the use of nano-heterostructures consisting of topological insulator nanostructures grown on 2D materials. In this work, we report the synthesis of TI/graphene nano-heterostructures based on $Bi_2Te_3$ and $Sb_2Te_3$ nanoplatelets (NPs) grown on single-layer graphene. Various techniques were used to characterize this system in terms of morphology, thickness, composition, and crystal quality. We found that most of the obtained NPs are mainly < 20 [nm] thick with thickness-dependent crystal quality, observed by Raman measurements. Thinner NPs (1 or 2 QL) tend to replicate the topography of the underlying SLG, according to roughness analysis, and observed buckling features. Finally, we show preliminary studies of their band structure obtained by LT-STM (STS) and by DFT. We observe a highly negative doping which can be attributed to the presence of defects.

Electronic Supplementary Information (ESI): Optical microscopy of synthesis with different growth parameters, optical microscopy and Raman spectroscopy of Te rods, optical microscopy images of $Sb_2Te_3$/SLG nano-heterostructure, close-ups of AFM images with their corresponding height profiles, graph of RMS of NPs and Graphene as a function of height, graph of <G(r)>, height profile of buckling-type pattern and Density Functional Theory calculations.

## Introduction

Topological insulators (TIs) are a new class of quantum material with a bulk gap and topologically protected massless Dirac surface states [1]. Since their prediction [2] and experimental realization [3], TIs have attracted great interest due to their potential for future applications in novel technologies such as spintronics [4], quantum computing [5] and low-dissipation electronic devices [6] as well as their potential to investigate fundamental phenomena such as Majorana fermions [7], proximity-induced superconductivity [8], and quantum anomalous Hall effect [9].

To exploit these unprecedented properties and phenomena, access to topological surface states is required. However, one of the most important challenges to overcome in the field of TI research is the presence of natural defects in the material, which produce intrinsic doping that leads to a contribution of the bulk states to the material's conductivity. Various routes have been used to diminish the bulk conductance, for example, external doping [10–12] has been extensively used to compensate for bulk doping through a high residual dopant concentration but can lead to degrading electron carrier mobility [5]. The use of gating [13,14], different substrates [15,16], strain engineering [17–19] and thickness control [20] have also been proposed to manipulate the electronic structure of the material.

Among these methods, both strain engineering and thickness control are possible to obtain through the use of nano-heterostructures consisting of topological insulator nanostructures grown on 2D materials. The use of nanostructured TI has been proposed as an alternative to suppress bulk conductivity, due to their large surface-to-bulk ratio [21,22] and the possibility of controlling the topological surface state via the quantum confinement effect in the < 30 [nm] thickness range [23,24]. Moreover, these nanostructures can be grown onto a variety of other materials, thus creating strained TI systems. The introduction of strain in TIs has been reported to tune the Dirac surface states [25], and to induce superconductivity [26] and van Hove singularities [27].

In addition, new emerging phenomena have been observed in various TI heterostructures systems such as induced magnetoresistance [28] and multiferroic properties [29] in $Sb_2Te_3$/GeTe heterostructures and an enhancement of optical sensitivity in TI/Ferromagnet heterostructures [30]. In the case of TI/graphene heterostructures, different proximity effects have been reported such as Rashba splitting and Dirac point ($E_D$) shift in the TI [31], the appearance of heavy Dirac fermions [32] and plasmonic excitations [33] across the heterojunction, and giant spin-orbit coupling in graphene, resulting in a new 2D system with non-trivial spin texture and high electron mobility [34–36].

To this day, the majority of studies on TI/graphene heterostructures have been made on thin films grown by MBE [15,37]. This method of synthesis, while useful to obtain high-quality crystals, is expensive and slow, making it less accessible and scalable. Other methods to obtain these crystals include (i) mechanical exfoliation, which lacks control of the crystals' size and thickness, and (ii) solvothermal synthesis, which is a low-cost alternative, but the purity of the materials is not comparable to other techniques [38]. In contrast, chemical vapor deposition (CVD) appears as a powerful and low-cost bottom-up synthesis method that allows to achieve high-quality TI nanostructures in a variety of substrates compatible with device applications [5]. In this work, we focus on the growth and characterization of $Bi_2Te_3$ and $Sb_2Te_3$ nanostructures grown by CVD on single-layer graphene (SLG) to form nano-heterostructures. TIs synthesis parameters are tuned to keep TI nanostructures thickness mainly below the range (< 30 [nm]) where the proximity and confinement effects have been reported [23,24].

$Bi_2Te_3$ and $Sb_2Te_3$ are chalcogenide materials that have a rhombohedral crystalline structure with a lattice constant of 0.4395 [nm] and 0.4262 [nm] correspondingly. Their structure (Fig. 1c) consists of five atomic sheets, called quintuple layers (QLs), with a thickness of 1 [nm] that are weakly bonded to each other by van der Waals (vdW) interactions. On the other side, graphene is a single layer of carbon atoms, arranged in a hexagonal lattice with a lattice constant of 0.246 [nm]. These materials bond to each other via interlayer vdW interactions, thus adding strain to the layers due to the initial lattice mismatch of the materials [19]. In the system presented here, $Bi_2Te_3$ and $Sb_2Te_3$ nanostructures can be epitaxially grown on graphene due to their small lattice mismatch of 2.7% for $Bi_2Te_3$ [39] and 4% for $Sb_2Te_3$ [40] thus resulting in a small in-plane strain.

In general, the electronic structure of TIs has been mostly determined by angle-resolved photoemission spectroscopy (ARPES) performed on MBE-grown thin films [41] or ultra-high vacuum (UHV) cleaved crystals [42]. Since

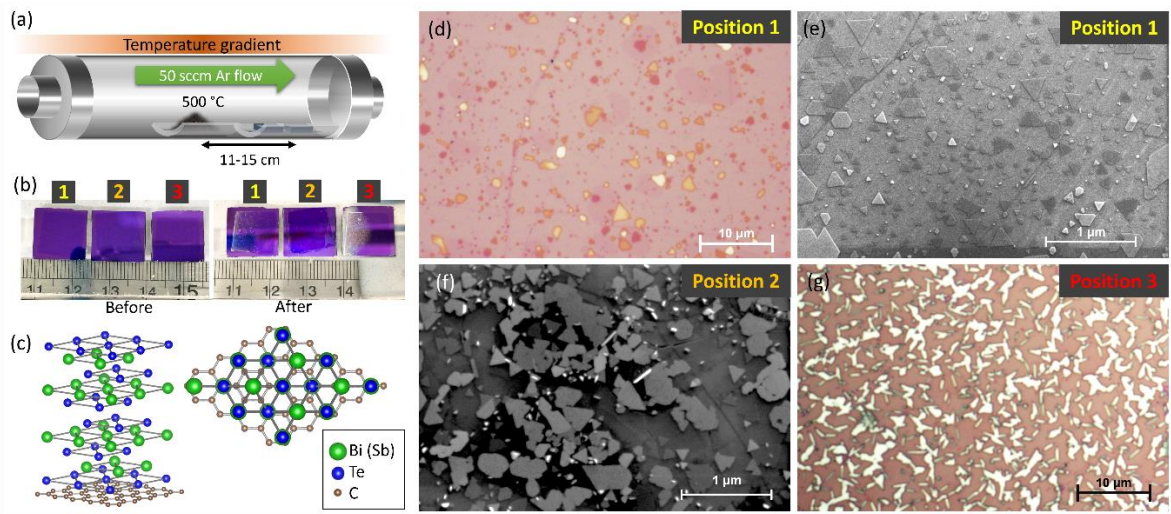

**Fig. 1 (a)** Schematic drawing of vapor-solid CVD deposition method. **(b)** Substrates before and after synthesis of $Bi_2Te_3$/SLG nano-heterostructure. **(c)** Crystal structure of $Bi_2Te_3$/SLG ($Sb_2Te_3$/SLG) nano-heterostructure. **(d)-(g)** Optical microscopy and SEM images corresponding to the substrates shown in **(b)**.

ARPES signal cannot obtain local measurements due to beam size limitations, this restricts the access to high spatial resolution of TI band structure. In contrast, scanning tunneling spectroscopy (STS) allows access to local phenomena and has been used in the past to observe the absence of backscattering from nonmagnetic impurities of TI surface states [43,44]. However, few measurements of the local density of states (LDOS) of TI nanostructures have been carried out using STS, once again being focused on thin films [15] or UHV-cleaved crystals [44].

Considering this, we explore the structural and electronic properties of TI/SLG nano-heterostructures based on thin $Bi_2Te_3$ and $Sb_2Te_3$ nanoplatelets (NPs), grown on single-layer graphene. Synthesis parameters were tuned during growth to obtain TI nanostructures of <30 [nm] thicknesses to favor the appearance of surface states and interaction with SLG substrate. Various techniques were used to characterize these nano-heterostructures in terms of morphology, thickness, composition, and crystal quality, such as Raman Spectroscopy, SEM, EDS, AFM and LT-STM, together with preliminary theoretical and experimental studies of their band structure, obtained by DFT and scanning tunneling spectroscopy (STS).

## Materials and Methods

Before TI NPs synthesis, single-layer graphene is transferred onto $SiO_2$ substrates (Silicon Prime Wafer, Oxide Thickness: 285 nm, N-doped) by a PMMA-assisted method [45]. $Bi_2Te_3$ and $Sb_2Te_3$ NPs were synthesized by a catalyst-free vapor transport and deposition process, which is described in Fig. 1a. The corresponding powders (Bismuth(III) telluride, vacuum deposition grade, 99.999% (metals basis – Alfa Aesar) and antimony (III) telluride, 99.999% (metals basis – Alfa Aesar) are placed on a quartz plate in the middle of one zone of the two-zone furnace (TFM2-1200 Across International). The substrates are placed 11-15 [cm] downstream of the powder source. Other variations of synthesis parameters were optimized before (See Fig. S1).

Synthesis is carried out for 5 [min] at a temperature of 500 °C and with an Ar flow of 50 [sccm]. Once the growth is complete, it is allowed to cool naturally (~3.5 [°C/min]), maintaining the flow of Ar and the system pressure (~0.3 [torr]). The TI NPs/SLG samples obtained were characterized by optical microscopy, SEM and AFM to confirm the expected morphology and structure, together with Raman and EDS spectroscopy to obtain information on the crystalline quality and its composition. Finally, nano-heterostructures were characterized by LT-STM (STS) to obtain information on their local electronic properties, which were compared to DFT calculations.

## Results

Regarding the effect of the growth parameters on the resulting NPs grown on SLG, it was observed that changes in the temperature of the substrate, the synthesis time, and the Ar gas flow affected the morphology and density of the grown nanostructures (Fig. 1d-g). These preliminary syntheses allowed us to obtain the growth parameters required for the target NP thickness.

As shown in Fig. 1b, three distinct positions for the different substrate temperatures ($T_S$) were identified, starting with position 1 at the highest $T_S$ and position 3 at the lowest $T_S$. After the growth process, differences between samples are evident to the naked eye (Fig. 1b). Optical microscopy and SEM images (Fig. 1d-g) show the difference between the nanostructures obtained at the different $T_S$ positions. Samples at position 1 (Fig. 1d and 1e) show the targeted nano-heterostructure of NPs that have a hexagonal or triangular shape, characteristic of the crystalline orientation of this material. These NPs have lateral dimensions between 0.1-2 [μm], with a high growth density without percolation.

The other two positions, located only 1-2 [cm] away from the sample at position 1 (Fig. 1f-g) obtain NPs with Te rods (position 2) or only Te rods (position 3), as confirmed with Raman spectroscopy (Fig. S2) that do not evidence the expected peaks of bismuth telluride and instead presented the peaks for crystalline tellurium [46,47]. These differences are also observed in $Sb_2Te_3$/SLG nano-heterostructures (Fig. S3). This very small distance range in which ideal samples are synthesized demonstrates the importance of correctly tuning the studied growth parameters to minimize the waste of resources.

To study the composition of the NPs obtained, EDS and Raman spectroscopy measurements were performed. Fig. 2a shows an SEM image with EDS maps of Bi and Te signals overlapped and each element separately, confirming the homogeneity in the composition of the nanostructures. Fig. 2b shows the Raman spectra of three

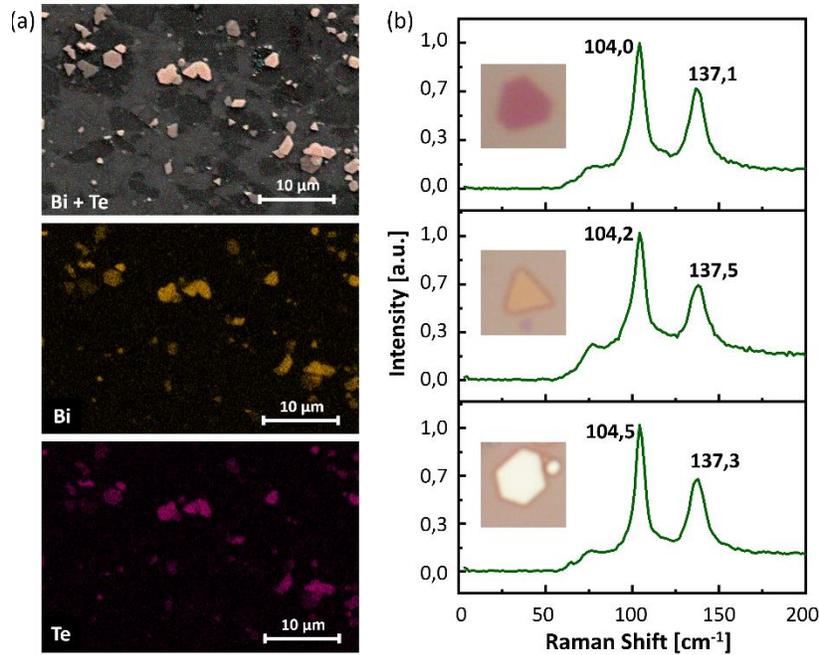

**Fig. 2 (a)** EDS maps of $Bi_2Te_3$/SLG nano-heterostructure, where the upper image shows a SEM image with both elements' signals combined, these signals are shown in the images below. **(b)** Raman spectra of individual $Bi_2Te_3$ NPs of different thicknesses, where the upper one is the thinnest and the bottom one is the thickest

NPs of different thicknesses, starting from the thinnest (purple) to the thickest (yellow), these relative thicknesses being estimated according to the coloration of the NPs [48].

In the Raman spectra, the $E^2_g$ (~104 [cm$^{-1}$]) and $A^2_{1g}$ (~137 [cm$^{-1}$]) peaks characteristic of the bulk material [49] are marked, indicating a high crystalline quality along with the small FWHM of the thick NPs (See Table S1). An increase in the FWHM is observed, when decreasing the thickness of the NPs, suggesting the presence of defects or impurities in the thinner NPs. Also, in the thinnest NP, the appearance of the $A^1_u$ peak (~119 [cm$^{-1}$]) is observed, which has been reported in thin sheets and attributed to a break in the symmetry of the crystal in the third dimension due to the limited thickness [50]. Regarding the center of the peaks, a small variation (~0.5 [cm$^{-1}$]) is observed between the thickest and thinnest NPs, indicating little to no strain caused by the substrate in which it has been synthesized [51].

Height profile analysis of AFM images, as seen in Fig. 3a, show that most of the NPs in the $Bi_2Te_3$/SLG nano-heterostructure were within the target thickness (< 30 nm). In addition, the growth direction of the NPs is indicated (with green arrows), a characteristic direction of van der Waals growth, confirming this type of growth. AFM measurements for the $Sb_2Te_3$/SLG nano-heterostructure show similar results (Fig. 3b).

Fig. 3a-b show NPs of different heights and wrinkles in graphene (see Fig. S4 for close-ups of these NPs and graphene wrinkles with their corresponding height profiles). Such wrinkles in graphene, intrinsic to the graphene transfer process, seem to perform as a barrier that stops the epitaxial growth of NPs, as evidenced by various NPs which present a truncated flat side where the NP meets the wrinkle, thus showing the importance of a flat substrate to favor epitaxial growth over vertical growth, and thus, to obtain thin NPs.

Roughness analyses were performed on the nano-heterostructures, which can be seen in Fig. 3c. This graph shows the dispersion and average value of the roughness of the graphene, in red, and of the NPs, in blue for $Bi_2Te_3$ and green for $Sb_2Te_3$. From this data, a trend between the thickness of the NPs and their roughness was obtained, where the thinnest NPs present higher roughness than the thicker NPs. This increase in roughness for thinner NPs can be explained by the topographic characteristics of the underlying graphene affecting the growth of the NPs. Presumably, the graphene roughness replicates on the surface of the NPs, whereas thicker NPs are less affected by the substrate and thus their roughness homogenizes and tends to a bulk value. It is observed that thinner NPs present a roughness similar to the one obtained for graphene and that the dispersion of the data

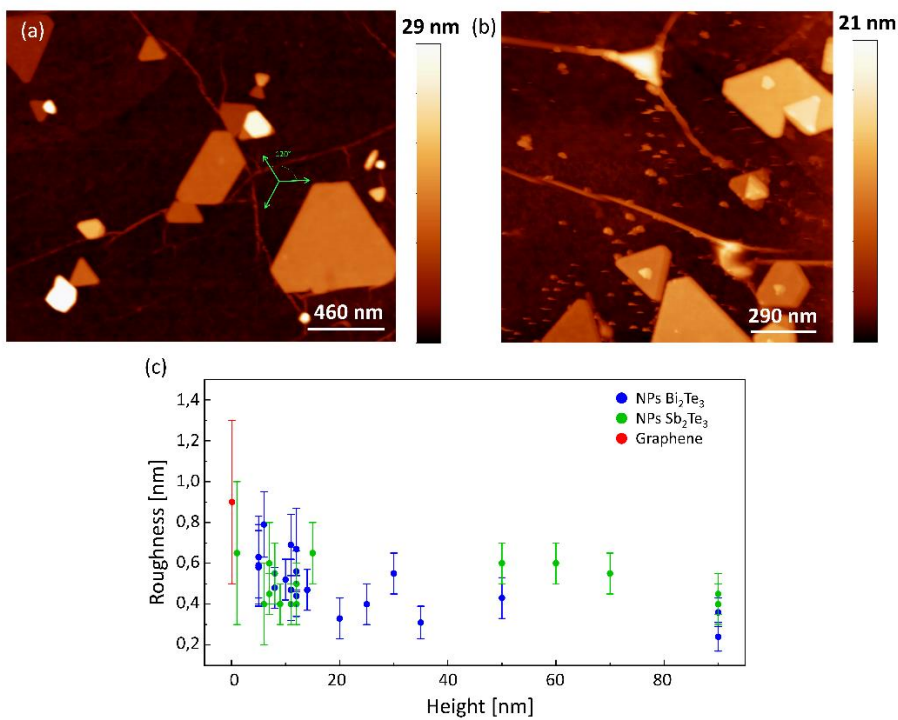

**Fig. 3 (a)** AFM image of $Bi_2Te_3$/SLG nano-heterostructure. The green arrows show the characteristic direction growth of van der Waals epitaxy. **(b)** AFM image of $Sb_2Te_3$/SLG nano-heterostructure. **(c)** Graph of roughness of NPs and Graphene as a function of height. Various NPs and different Graphene zones of one sample were used. Error bars represent the confidence interval in roughness graphs. A tendency of roughness to decrease with NP height can be observed.

decreases as the thickness of the NPs increases, which supports this interpretation. It was also observed that the root mean square (RMS) values follow a similar pattern (Fig. S5).

Atomic-resolved topographic images of SLG were obtained using STM, as shown in Fig. 4a. Fig. 4b shows a zoom where the characteristic honeycomb hexagonal lattice can be seen. In Fig. 4c the corresponding fast Fourier transform (FFT) is observed, with which a lattice parameter of (1.41±0.07) [Å] was obtained, an expected value for single-layer graphene.

STM images show NPs as thin as 2 QL, like the one shown in Fig. 4d. Atomic resolved images of the NP (Fig. 4e) and FFT analysis (Fig. 4f), confirm a lattice parameter of (4.52±0.15) [Å], consistent with $Bi_2Te_3$ structure.

Regarding their band structure, different studies have shown that these materials present band structures affected by their bulk states. In the case of $Bi_2Te_3$, the Dirac point is buried in the bulk valence band, resulting in a n-doped material [52,53], while in the case of $Sb_2Te_3$, the Dirac point is located in the bulk band gap, resulting in a

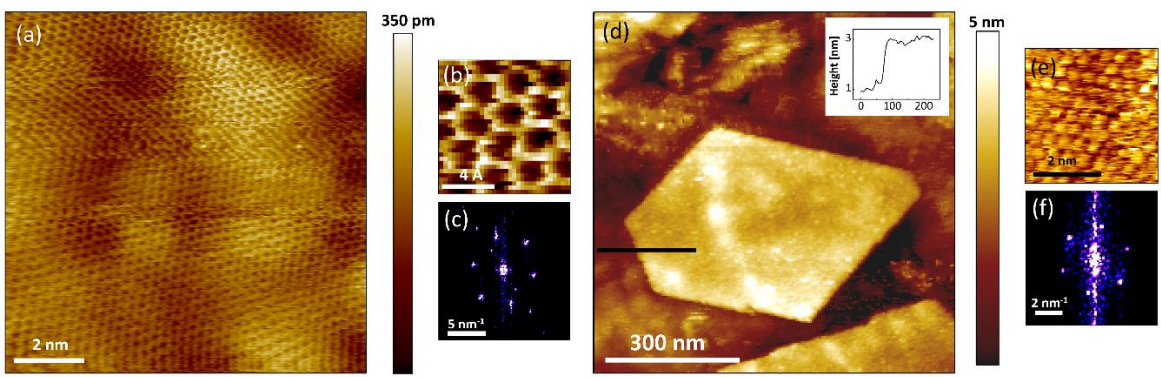

**Fig. 4 (a)** STM image of single-layer graphene transferred on $SiO_2$ with **(b)** close-up showing its characteristic honeycomb hexagonal lattice and **(c)** corresponding FFT. I=30 [pA]; $V_B$=0,8 [V] **(d)** STM image of 2 QL (height profile in inset) $Bi_2Te_3$ NP with **(e)** close-up showing its lattice and **(f)** corresponding FFT. I=10 [pA]; $V_B$=0,5 [V], T=70 [K].

p-doped material [15,53]. STS measurements on $Bi_2Te_3$ and $Sb_2Te_3$ nano-heterostructures show negative doping for both materials. Fig. 5a shows representative dI/dV spectra with a spatial resolution of 0.16 [nm]. These spectra are taken along a line in a 10 QL $Bi_2Te_3$ NP where a yellow dotted line marks the position of the average Dirac point. The Dirac points obtained by each spectrum can be seen in Fig. 5b, where the average value is marked with a dotted line. Fig. 5c and 5d show the same analysis for a 5 QL $Sb_2Te_3$ NP.

The Dirac point of $Bi_2Te_3$ NPs is obtained by finding the intersection between the linear surface states and 0 LDOS. A mean $E_D$ value ~1.1 [V] below the Fermi level is found, making them negative-doped. This doping is an expected result due to the effect of the intrinsic defects in this material [52]. It is known that Te-on-Bi antisite defects ($Te_{Bi}$) give rise to negative doping in $Bi_2Te_3$ [52], thus, this kind of defects could contribute to the high negative doping of the nano-heterostructure. We also performed DFT calculations for a 10 QL NP grown on SLG (See Fig. S6). Here, we obtain a large difference in the location of the Dirac point between the experimental and calculated values, once again, this difference can be attributed to the intrinsic defects obtained during synthesis, as these defects are not present in our DFT calculations. Compared to experimental results reported by another study of $Bi_2Te_3$ NPs we also observe a difference in the location of the Dirac point; however, these are grown on $SiO_2$ [53] and HOPG [20] instead of SLG. Additionally, a relation between the thickness of the NPs and their doping has been previously reported [20]. Here, thinner NPs are more negatively doped than thicker NPs and this could explain the high negativity found for the 10 QL NP.

In addition, as seen in Fig. 5b, variations of the $E_D$ were obtained, which indicates inhomogeneities in the composition of one individual NP. These inhomogeneities could coincide with the presence of defects and impurities observed by Raman spectroscopy in the thinnest NPs.

For $Sb_2Te_3$ NPs (Fig. 5c), the dI/dV spectra show a mean bulk band gap value of 0.26 [V] which is consistent with previous reports of $Sb_2Te_3$ *thin films* grown on graphene by MBE [15] but bigger than our DFT calculations (Figure

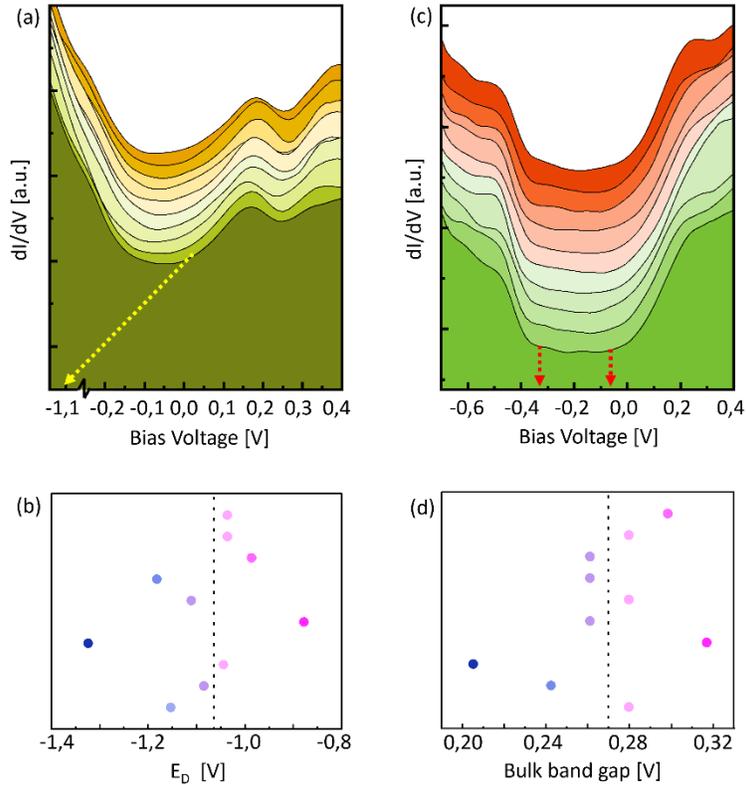

**Fig. 5 (a)** Measurements of dI/dV along a line in a $Bi_2Te_3$ NP of 10 QL. The yellow dotted line shows the mean $E_D$ value. **(b)** Position of the Dirac point for each spectrum displayed in **(a)** with the mean value marked with a black dotted line. **(c)** Measurements of dI/dV along a line in a $Sb_2Te_3$ NP of 5 QL. The yellow dotted lines show the mean bulk band gap value. **(d)** Bulk band gap for each spectrum is displayed in **(c)** with the mean value marked with a black dotted line.

Fig. S7). This could be explained by an incorrect description of the band gap of the material and/or the presence of defects. For this material, the Dirac point lies within this gap, so these results suggest a negative doping due to the position of the bulk band gap. It is known that this material has intrinsic positive doping, however, as mentioned before, a shift in the $E_D$ towards negative doping as the thickness of NPs decreases has been observed on $Sb_2Te_3$ *thin films* [15]. Since the $Sb_2Te_3$ NP presented here is only 5 QL thick, this negative doping is consistent with these previous reports.

Finally, a buckling pattern was observed on the surface of a 1 QL $Sb_2Te_3$ NP (Fig. 6a). The average spatial correlation function <G(r)> (Fig. S8a) and the angle-dependent spatial correlation function <$G_θ(r)$> [54] (Fig. 6b) were computed for the topographic image shown in Fig. 6a. <$G_θ(r)$> and reflects the spatial pattern observed in the topographic image and clearly shows that there is a particular spatial pattern with a characteristic scale length. In Fig. 6b, arcs of intensity that repeat with a fixed periodicity are marked, these imply a pattern of phase separation of low and high topographic regions over large length scales. These arcs of intensity are consistent with our real space phase separation of a topographic striped pattern. The functional form of these arcs for a system with stripes separated by a distance d=~8.75 [nm] and running along an angle α from the horizontal, measured at an angle θ is given by Nd/cos((α-90)-θ).

As it was mentioned before, the topographic characteristics of the underlying graphene can be replicated on the surface of the NPs. It has been reported that due to thermal cycling, a buckling effect can occur on graphene triggered by a compressive strain [55]. The nano-heterostructure samples presented here undergo a thermal treatment before LT-STM measurements as well, so it is expected that NPs reproduce a similar buckling topography. This kind of strained topography has been reported to induce a periodically modulated pseudo-magnetic field [55], however, there are no previous reports for this type of buckling on TI nanostructures. Although the height modulation of the observed buckling pattern is around 0.6 [nm] (See Fig. S8b), the mean surface roughness is ~0.18 [nm], a value similar to the one found in buckled graphene superlattices [55].

As a preliminary approach, we obtained representative point spectra for this NP (Fig. 6c) and found similar results as seen on $Sb_2Te_3$ 4 QL *thin films* grown by MBE on graphene [15]. They obtained a shift in the $E_D$ towards negative doping as the thickness of NPs decreases. This is consistent with our results, as their thinnest film, of 4 QL, had a $E_D$ of -100 [mV], while our 1 QL NP has a $E_D$ of -150 [mV].

## Conclusions

We explore the properties of TI/SLG nano-heterostructures based on <20 [nm] thick $Bi_2Te_3$ and $Sb_2Te_3$ NPs, grown on monolayer graphene by CVD. We observed the topographic influence of SLG substrate on roughness and topography of grown TI NPs. Crystal quality decreases with NPs thickness and the appearance of peaks linked to a break in the symmetry of the system due to size limitations, while the thicker NPs showed the expected bulk spectra.

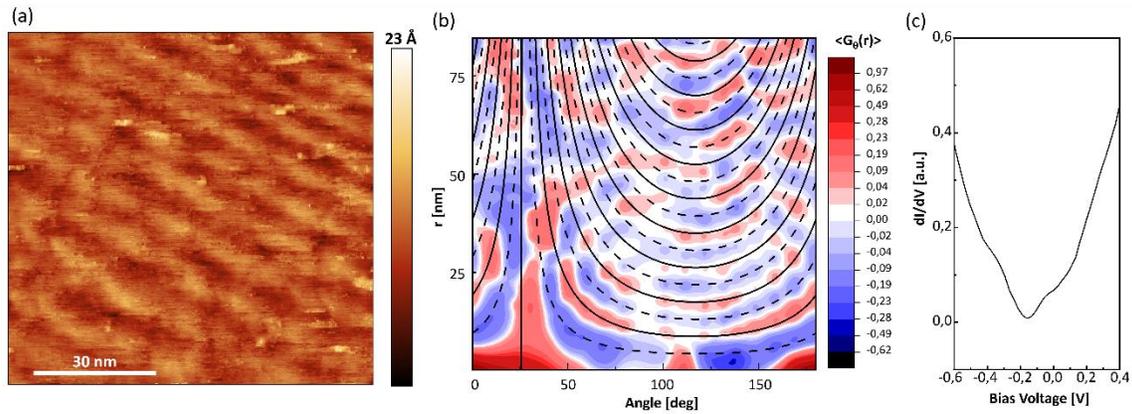

**Fig. 6 (a)** STM image of the surface of a $Sb_2Te_3$ NP where a striped pattern can be observed. **(b)** Angle-dependent spatial correlation graph of the image in **(a)**. Arcs of intensity given by Nd/cos((α-90)-θ) are marked with a continuous black line for maxima and a dashed black line for minima. **(c)** dI/dV representative spectra of NP shown in **(a)**.

The STS measurements showed a highly negative doping for both nano-heterostructures. This doping is probably connected to the presence of Te$_{Bi}$ defects on the composition of the NPs. Since a diminution of crystal quality on thinner NPs was observed, this is a possible origin of doping. The NPs also showed $E_D$ spatial inhomogeneities within a single NP, which is also consistent with the presence of defects.

Further STS studies on this nano-heterostructure system will provide insight about a possible NP thickness dependence of $E_D$ and the effect of the buckling-type pattern on the LDOS of the nano-heterostructure.

Control of the morphology of the graphene and thickness of NPs in this kind of nano-heterostructure could open a new route to generate buckling in TI materials that will lead to emergent electronic properties, such as pseudo-magnetic fields reported on graphene-based strained structures.

## Author Contributions

The manuscript was written through contributions of all authors. All authors have given approval to the final version of the manuscript.

## Conflicts of interest

There are no conflicts to declare.

## Acknowledgements

The authors acknowledge the financial support from Fondo Nacional de Investigación Científica y Tecnológica, grants 1220715 and 1231487 and Agencia Nacional de Investigación y Desarrollo, grant Anillo ACT192023 and Beca de Magíster Nacional folio 22201487. The authors thank projects FONDEQUIP EQM190177, FONDEQUIP EQM190179, FONDEQUIP EQM200085 and ANID PIA/APOYO AFB220004. This work was also partially supported by the Center for the Development of Nanoscience and Nanotechnology CEDENNA AFB220001 and by the supercomputing infrastructure of the NLHPC (ECM-02)

## Notes and references

1 C. L. Kane and E. J. Mele, *Phys Rev Lett*, 2005, **95**, 146802.

2 B. A. Bernevig, T. L. Hughes and S.-C. Zhang, *Science (1979)*, 2006, **314**, 1757–1761.

3 M. König, S. Wiedmann, C. Brüne, A. Roth, H. Buhmann, L. W. Molenkamp, X.-L. Qi and S.-C. Zhang, *Science (1979)*, 2007, **318**, 766–770.

4 Y. Tokura, K. Yasuda and A. Tsukazaki, *Nature Reviews Physics*, 2019, **1**, 126–143.

5 P. Liu, J. R. Williams and J. J. Cha, *Nat Rev Mater*, 2019, **4**, 479–496.

6 M. Nadeem, A. R. Hamilton, M. S. Fuhrer and X. Wang, *Small*, 2020, **16**, 1–13.

7 M. Z. Hasan and C. L. Kane, *Rev Mod Phys*, 2010, **82**, 3045–3067.

8 L. He, X. Kou and K. L. Wang, *physica status solidi (RRL) – Rapid Research Letters*, 2013, **7**, 50–63.

9 K. He, Y. Wang and Q.-K. Xue, *Annu Rev Condens Matter Phys*, 2018, **9**, 329–344.

10 H. Li, Y. R. Song, M.-Y. Yao, F. Zhu, C. Liu, C. L. Gao, J.-F. Jia, D. Qian, X. Yao, Y. J. Shi and D. Wu, *J Appl Phys*, 2013, **113**, 043926.

11 Z. Ren, A. A. Taskin, S. Sasaki, K. Segawa and Y. Ando, *Phys Rev B*, 2012, **85**, 155301.

12 Y. Tanaka, K. Nakayama, S. Souma, T. Sato, N. Xu, P. Zhang, P. Richard, H. Ding, Y. Suzuki, P. Das, K. Kadowaki and T. Takahashi, *Phys Rev B*, 2012, **85**, 125111.

13 F. Xiu, L. He, Y. Wang, L. Cheng, L.-T. Chang, M. Lang, G. Huang, X. Kou, Y. Zhou, X. Jiang, Z. Chen, J. Zou, A. Shailos and K. L. Wang, *Nat Nanotechnol*, 2011, **6**, 216–221.

14 J. Yao, J. Shao, Y. Wang, Z. Zhao and G. Yang, *Nanoscale*, 2015, **7**, 12535–12541.

15 Y. Jiang, Y. Y. Sun, M. Chen, Y. Wang, Z. Li, C. Song, K. He, L. Wang, X. Chen, Q.-K. Xue, X. Ma and S. B. Zhang, *Phys Rev Lett*, 2012, **108**, 066809.

16 A. Zalic, T. Dvir and H. Steinberg, *Phys Rev B*, 2017, **96**, 075104.

17 Y. Liu, Y. Y. Li, S. Rajput, D. Gilks, L. Lari, P. L. Galindo, M. Weinert, V. K. Lazarov and L. Li, *Nat Phys*, 2014, **10**, 294–299.

18 C. Schindler, C. Wiegand, J. Sichau, L. Tiemann, K. Nielsch, R. Zierold and R. H. Blick, *Appl Phys Lett*, 2017, **111**, 171601.

19 P. A. Vermeulen, J. Mulder, J. Momand and B. J. Kooi, *Nanoscale*, 2018, **10**, 1474–1480.

20 C. Parra, T. H. Rodrigues da Cunha, A. W. Contryman, D. Kong, F. Montero-Silva, P. H. Rezende Gonçalves, D. D. Dos Reis, P. Giraldo-Gallo, R. Segura, F. Olivares, F. Niestemski, Y. Cui, R. Magalhaes-Paniago and H. C. Manoharan, *Nano Lett*, 2017, **17**, 97–103.

21 Y. S. Kim, M. Brahlek, N. Bansal, E. Edrey, G. A. Kapilevich, K. Iida, M. Tanimura, Y. Horibe, S.-W. Cheong and S. Oh, *Phys Rev B*, 2011, **84**, 073109.

22 B. Xia, P. Ren, A. Sulaev, P. Liu, S.-Q. Shen and L. Wang, *Phys Rev B*, 2013, **87**, 085442.

23 S. S. Hong, D. Kong and Y. Cui, *MRS Bull*, 2014, **39**, 873–879.

24 R. Giraud and J. Dufouleur, *physica status solidi (b)*, 2021, **258**, 2000066.

25 D. Flötotto, Y. Bai, Y.-H. Chan, P. Chen, X. Wang, P. Rossi, C.-Z. Xu, C. Zhang, J. A. Hlevyack, J. D. Denlinger, H. Hong, M.-Y. Chou, E. J. Mittemeijer, J. N. Eckstein and T.-C. Chiang, *Nano Lett*, 2018, **18**, 5628–5632.

26 S. Charpentier, L. Galletti, G. Kunakova, R. Arpaia, Y. Song, R. Baghdadi, S. M. Wang, A. Kalaboukhov, E. Olsson, F. Tafuri, D. Golubev, J. Linder, T. Bauch and F. Lombardi, *Nat Commun*, 2017, **8**, 2019.

27 T.-H. Kim, K. Jeong, B. C. Park, H. Choi, S. H. Park, S. Jung, J. Park, K.-H. Jeong, J. W. Kim, J. H. Kim and M.-H. Cho, *Nanoscale*, 2016, **8**, 741–751.

28 J. Tominaga, Y. Saito, K. Mitrofanov, N. Inoue, P. Fons, A. V. Kolobov, H. Nakamura and N. Miyata, *Adv Funct Mater*, 2017, **27**, 1702243.

29 J. Tominaga, A. V Kolobov, P. J. Fons, X. Wang, Y. Saito, T. Nakano, M. Hase, S. Murakami, J. Herfort and Y. Takagaki, *Sci Technol Adv Mater*, 2015, **16**, 014402.

30 X. Li, Y. G. Semenov and K. W. Kim, *Appl Phys Lett*, 2014, **104**, 061116.

31 H.-D. Song, D. Sheng, A.-Q. Wang, J.-G. Li, D.-P. Yu and Z.-M. Liao, *Chinese Physics B*, 2017, **26**, 037301.

32 W. Cao, R.-X. Zhang, P. Tang, G. Yang, J. Sofo, W. Duan and C.-X. Liu, *2d Mater*, 2016, **3**, 034006.

33 Y. Liu, T. Wei, P. Cui, X. Li and Z. Zhang, *Phys Rev B*, 2022, **105**, 195150.

34 J. Zhang, C. Triola and E. Rossi, *Phys Rev Lett*, 2014, **112**, 096802.

35 K.-H. Jin and S.-H. Jhi, *Phys Rev B*, 2013, **87**, 075442.

36 K. Zollner and J. Fabian, *physica status solidi (b)*, 2021, **258**, 2000081.

37 Y. Yin, G. Wang, C. Liu, H. Huang, J. Chen, J. Liu, D. Guan, S. Wang, Y. Li, C. Liu, H. Zheng and J. Jia, *Nano Res*, 2022, **15**, 1115–1119.

38 W. Tian, W. Yu, J. Shi and Y. Wang, *Materials*, 2017, **10**, 814.

39  H. Qiao, J. Yuan, Z. Xu, C. Chen, S. Lin, Y. Wang, J. Song, Y. Liu, Q. Khan, H. Y. Hoh, C.-X. Pan, S. Li and Q. Bao, *ACS Nano*, 2015, **9**, 1886–1894.

40  S. Singh, S. Kim, W. Jeon, K. P. Dhakal, J. Kim and S. Baik, *Carbon N Y*, 2019, **153**, 164–172.

41  K. Zhang, H. Pan, Z. Wei, M. Zhang, F. Song, X. Wang and R. Zhang, *Chinese Physics B*, 2017, **26**, 096101.

42  D. Kong, W. Dang, J. J. Cha, H. Li, S. Meister, H. Peng, Z. Liu and Y. Cui, *Nano Lett*, 2010, **10**, 2245–2250.

43  P. Ngabonziza, M. P. Stehno, G. Koster and A. Brinkman, in *In-situ Characterization Techniques for Nanomaterials*, Springer Berlin Heidelberg, Berlin, Heidelberg, 2018, pp. 223–250.

44  Z. Alpichshev, J. G. Analytis, J.-H. Chu, I. R. Fisher, Y. L. Chen, Z. X. Shen, A. Fang and A. Kapitulnik, *Phys Rev Lett*, 2010, **104**, 016401.

45  C. Parra, J. Aristizabal, B. Arce, F. Montero-Silva, S. Lascano, R. Henriquez, P. Lazcano, P. Giraldo-Gallo, C. Ramírez, T. Henrique Rodrigues da Cunha and A. Barrera de Brito, *Metals (Basel)*, 2021, **11**, 147.

46  S. Khatun, A. Banerjee and A. J. Pal, *Nanoscale*, 2019, **11**, 3591–3598.

47  Y. Du, G. Qiu, Y. Wang, M. Si, X. Xu, W. Wu and P. D. Ye, *Nano Lett*, 2017, **17**, 3965–3973.

48  F. Yang, M. Sendova, R. B. Jacobs-Gedrim, E. S. Song, A. Green, P. Thiesen, A. Diebold and B. Yu, *AIP Adv*, 2017, **7**, 015114.

49  K. M. F. Shahil, M. Z. Hossain, D. Teweldebrhan and A. A. Balandin, *Appl Phys Lett*, 2010, **96**, 153103.

50  K. M. F. Shahil, M. Z. Hossain, V. Goyal and A. A. Balandin, *J Appl Phys*, 2012, **111**, 054305.

51  W. Dang, H. Peng, H. Li, P. Wang and Z. Liu, *Nano Lett*, 2010, **10**, 2870–2876.

52  G. Wang, X. Zhu, Y. Sun, Y. Li, T. Zhang, J. Wen, X. Chen, K. He, L. Wang, X. Ma, J. Jia, S. B. Zhang and Q. Xue, *Advanced Materials*, 2011, **23**, 2929–2932.

53  D. Kong, Y. Chen, J. J. Cha, Q. Zhang, J. G. Analytis, K. Lai, Z. Liu, S. S. Hong, K. J. Koski, S.-K. Mo, Z. Hussain, I. R. Fisher, Z.-X. Shen and Y. Cui, *Nat Nanotechnol*, 2011, **6**, 705–709.

54  P. Giraldo-Gallo, Y. Zhang, C. Parra, H. C. Manoharan, M. R. Beasley, T. H. Geballe, M. J. Kramer and I. R. Fisher, *Nat Commun*, 2015, **6**, 8231.

55  J. Mao, S. P. Milovanović, M. Anđelković, X. Lai, Y. Cao, K. Watanabe, T. Taniguchi, L. Covaci, F. M. Peeters, A. K. Geim, Y. Jiang and E. Y. Andrei, *Nature*, 2020, **584**, 215–220.